\documentclass[reprint,aps]{revtex4-1}
\usepackage{amsmath}
\usepackage{graphicx}
\usepackage{dcolumn}
\usepackage{bm}

\begin{document}

\title{Pinned modes in two-dimensional lossy lattices with local gain and
nonlinearity}
\author{Edwin Ding}
\address{Department of Mathematics and Physics, Azusa Pacific University, Box
7000, Azusa, CA 91702-7000, USA}

\author{A. Y. S. Tang and K. W. Chow}
\address{Department of Mechanical Engineering, University of Hong
Kong, Pokfulam Road, Hong Kong}

\author{Boris A. Malomed}
\address{Department of Physical Electronics, School of Electrical
Engineering, Faculty of Engineering, Tel Aviv University, Tel Aviv
69978, Israel}


\begin{abstract}
We introduce a system with one or two amplified nonlinear sites
(\textquotedblleft hot spots", HSs) embedded into a two-dimensional linear
lossy lattice. The system describes an array of evanescently coupled optical
or plasmonic waveguides, with gain applied to selected HS cores. The subject
of the analysis is discrete solitons pinned to the HSs. The shape of the
localized modes is found in quasi-analytical and numerical forms, using a
truncated lattice for the analytical consideration. Stability eigenvalues
are computed numerically, and the results are supplemented by direct
numerical simulations. In the case of self-focusing nonlinearity, the modes
pinned to a single HS are stable and unstable when the nonlinearity includes
the cubic loss and gain, respectively. If the nonlinearity is
self-defocusing, the \emph{unsaturated} cubic gain acting at the HS supports
\emph{stable} modes in a small parametric area, while weak cubic loss gives
rise to a \emph{bistability} of the discrete solitons. Symmetric and
antisymmetric modes pinned to a symmetric set of two HSs are considered as
well.
\end{abstract}

\maketitle


\section{Introduction}

Modes of fundamental significance to nonlinear optics~\cite{Rosanov,MT} and
plasmonics~\cite%
{plas1,plas3,plas4,plas5,plas6,plas7,plas8,plas9,Marini,Korea} are
dissipative spatial solitons that result from the simultaneous balance among
diffraction, self-focusing nonlinearity, loss, and compensating gain.
Stability is a crucially important issue in the theoretical analysis of
dissipative solitons. An obvious necessary condition for the stability of
localized modes is the stability of the zero background around them. The
basic complex Ginzburg-Landau (CGL) equation, which includes the
bandwidth-limited linear gain and nonlinear loss acting on a single field,
is unable to produce stable dissipative solitons, since the action of the
linear gain on the zero background makes it unstable. On the other hand,
dissipative solitons can be fully stabilized in systems of linearly coupled
CGL equations~\cite{wg1,wg2} modeling dual-core waveguides, with the linear
gain and loss acting in different cores~\cite%
{Marini,Dual1,Dual2,Dual3,Dual4,Dual5}, including the $\mathcal{PT}$%
-symmetric version of the system that features the balance between the gain
and loss~\cite{PT1,PT2}. Stable solitons can also be generated by a single
CGL equation with cubic gain \textquotedblleft sandwiched" between linear
and quintic losses, which may be realized in optics as a combination of
linear amplification and power-dependent absorption~\cite%
{CQ1,CQ2,CQ3,CQ4,CQ5,CQ6,cqgle1,cqgle2,cqgle3,cqgle4,cqgle5}.

Another method for creating stable localized modes makes use of linear gain
applied at a \textquotedblleft hot spot" (HS, i.e. a localized amplifying
region embedded into an ordinary lossy waveguide~\cite{HSexact,HS,Valery} or
a Bragg grating~\cite{Mak}). Models with multiple HSs~\cite%
{spotsExact1,spots1,spots2,spots3,spotsExact2}, and similar extended
amplifying structures~\cite{Zezyu1,Zezyu2}, have been studied as well. HSs
can be built by implanting gain-producing dopants into a narrow segment of
the waveguide~\cite{Kip}, or, alternatively, by focusing an external pump
beam at the designated position of the HS in a uniformly-doped waveguide. In
addition to models with the localized direct (phase-insensitive) gain,
systems including the localization of parametric gain was developed as well~%
\cite{param}.

Dissipative solitons can be stably pinned to the HS due to the balance
between the local gain and uniform loss in the bulk waveguide. For narrow
HSs modeled by the delta-functional distribution of the gain, solutions for
the pinned dissipative solitons are available in analytical form~\cite%
{HSexact,spotsExact1,spotsExact2}. Furthermore, models with mutually
balanced gain and loss applied in the form of $\delta $-functions at
separated points \cite{Stuttgart}, or at a single location, in the form of a
gain-loss dipole described by the derivative of the $\delta $-function \cite%
{Thawatchai}, make it possible to find solution for $\mathcal{PT}$-symmetric
solitons pinned to these points. Other one- and two-dimensional (1D and 2D)
HS-pinned modes, including stable vortices fed by the gain confined to an
annular-shaped area~\cite{2D1,2D2,2D3,2D4,2D6}, can be found numerically~%
\cite{HS,Valery,spots1,spots2,spots3}.

While dissipative solitons in uniform media are always unstable against the
blowup in the absence of the higher-order (quintic) nonlinear losses~\cite%
{Kramer,ml}, it is worthy to note a counter-intuitive result \cite{Valery}
demonstrating that \emph{stable} dissipative localized modes in uniform
linearly-lossy media may be supported by \emph{unsaturated} localized cubic
gain alone. Stable dissipative solitons were also predicted in a setting
that combines the uniformly-distributed linear gain in the $D$-dimensional
space and nonlinear loss growing from the center to periphery faster than $%
r^{D}$, where $r$ is the radial coordinate~\cite{Barcelona}.

The class of models with the localized gain includes lattice systems. In
Ref. \cite{we}, the 1D model was introduced for a linear lossy lattice with
a single or two amplified (active) sites embedded into it, which represent
the HSs in the discrete setting. It was assumed that the nonlinearity was
carried solely by the same active sites. This system, which may be
considered as a variety of discrete CGL equations \cite%
{discrCGL1,discrCGL2,discrCGL3,discrCGL4,discrCGL5,discrCGL6,discrCGL7,discrCGL8,discrCGL9,discrCGL10,discrCGL11}%
, can be implemented in the experiment using arrays of optical waveguides
\cite{review} or arrayed plasmonic waveguides ~\cite%
{plasmon-array1,plasmon-array2,plasmon-array3}. In particular, it suggests
possibilities for selective excitation of particular core(s) in the arrayed
waveguides, if the system is uniformly doped, but only the selected cores
are pumped by an external laser beam. In Ref. \cite{we}, discrete solitons
pinned to the HS in the lattice system were found in an analytical form,
similar to the soliton solutions available in the discrete linear Schr\"{o}%
dinger equation with embedded nonlinear elements~\cite{embed1,embed2,embed3}%
, and the stability of the localized models was investigated by means of
numerical methods.

The present work aims to extend the analysis for the 2D lattice system, with
one or two active nonlinear sites\ embedded into the linear lossy bulk
lattice. The experimental realization of such a 2D medium is also possible
in nonlinear optics, using waveguiding arrays permanently written in bulk
silica~\cite{Jena1,Jena2}. An essential distinction from the 1D counterpart~%
\cite{we} mentioned above is that 2D localized lattice modes cannot be found
analytically, even if only a single nonlinear site is embedded into the
linear matrix. Nevertheless, we demonstrate that semi-analytical solutions
can be obtained for truncated (finite-size) lattices.

The paper is organized as follows. The discrete 2D CGL equation is
introduced in Sec.~\ref{sec:gov}. Section~\ref{sec:trun} presents
semi-analytical results for the truncated lattices. Results of the
linear-stability analysis for the HS-pinned lattice solitons against small
perturbations are reported in Sec.~\ref{sec:stab}. In Sec.~\ref{sec:onset}
we extend the stability analysis, considering the crucially important issue
of the onset of the zero-solution instability. A brief consideration of the
double-HS system is presented in Sec.~\ref{sec:dual}. The paper is concluded
by Sec.~\ref{sec:con}.

\section{The model}

\label{sec:gov}

As said above, we consider the 2D generalization of the 1D\ lattice model
introduced in Ref. \cite{we}:
\begin{gather}
\frac{\mathrm{d}u_{m,n}}{\mathrm{d}z}=\frac{i}{2}\left(
u_{m-1,n}+u_{m+1,n}+u_{m,n-1}+u_{m,n+1}-4u_{m,n}\right)  \notag \\
-\gamma u_{m,n}+\left[ \left( \Gamma _{1}+i\Gamma _{2}\right) +\left(
iB-E\right) |u_{m,n}|^{2}\right] \delta _{m,0}\delta _{n,0}u_{m,n}\;,
\label{eq:gl}
\end{gather}%
where $m,n=0$, $\pm 1$, $\pm 2$, ... are discrete coordinates on the
lattice, $\delta _{m,0}$ and $\delta _{n,0}$ are the Kronecker's symbols,
and the coefficient of the linear coupling between adjacent cores is scaled
to unity. Further, $\gamma >0$ is the linear loss in the bulk lattice, $%
\Gamma _{1}>0$ and $\Gamma _{2}$ account for the linear gain and linear
potential applied at the HS site ($m=n=0$), while $B$ and $E$ account for
the Kerr nonlinearity and nonlinear loss/gain (for $E>0/E<0$) acting at the
HS.

In optics, the 2D discrete equation, as well as its 1D counterpart, can be
derived by means of well-known methods \cite%
{Chr,discrCGL1,discrCGL2,discrCGL3,discrCGL4,discrCGL5,discrCGL6,discrCGL7,discrCGL8,discrCGL9,discrCGL10,discrCGL11,review}%
. In the application to 2D arrays of plasmonic waveguides, which can be
built as a set of metallic nanowires embedded into a bulk dielectric \cite%
{plasmon-array1,plasmon-array2,plasmon-array3}, Eq. (\ref{eq:gl}) can be
derived in the adiabatic approximation, with the exciton field eliminated in
favor of the photonic one. It is relevant to mention that the well-known
\textit{staggering transformation} \cite{review}, $u_{m,n}(t)\equiv \left(
-1\right) ^{m+n}e^{-4it}\tilde{u}_{m,n}^{\ast }$, where the asterisk stands
for the complex conjugate, simultaneously reverses the signs of $\Gamma _{2}$
and $B$, thus rendering the self-focusing and defocusing signs of the
nonlinearity, which correspond to $B>0$ and $B<0$, respectively, mutually
convertible. This circumstance is essential, in particular, for modeling
arrays of plasmonic waveguides, where the intrinsic excitonic nonlinearity
is always self-repulsive. Below, we fix the signs of $\Gamma _{2}$ and $B$
by setting $\Gamma _{2}>0$, i.e. the corresponding linear potential at the
HS is \emph{attractive}, while $B$ may be positive (self-focusing), negative
(self-defocusing), or zero. Unless $B=0$, this coefficient is normalized to
be $B=\pm 1$. These two cases are considered separately below, along with
the case of $B=0$, when the nonlinearity is represented solely by the cubic
dissipation localized at the HS.

\section{The analysis for the truncated lattice}

\label{sec:trun}

\begin{figure}[t]
\begin{center}
\includegraphics[width = 80mm, keepaspectratio]{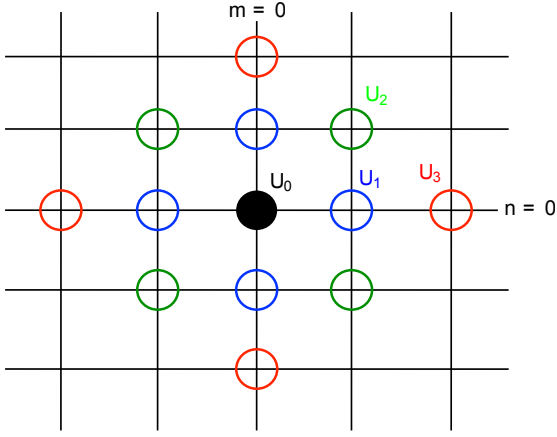}
\end{center}
\caption{(Color online) The sketch of the truncated square lattice, which
consists of 13 sites. The real amplitude, $U_{0}$ (black), and three complex
ones, $U_{1}$ (blue circles), $U_{2}$ (green circles), and $U_{3}$ (red
circles) are defined here as well. }
\label{sketch}
\end{figure}

We seek solutions to Eq. (\ref{eq:gl}) for stationary localized modes with
real propagation constant $k$ as
\begin{equation}
u_{m,n}(z)=U_{m,n}e^{ikz}\equiv \left( P_{m,n}+iQ_{m,n}\right) e^{ikz}.
\label{eq:pw}
\end{equation}%
Outside of the HS site, $m=n=0$, Eq. (\ref{eq:gl}) gives rise to the linear
stationary equation,%
\begin{equation}
2\left( k+2-i\gamma \right) U_{m,n}=U_{m-1,n}+U_{m+1,n}+U_{m,n-1}+U_{m,n+1},
\label{U}
\end{equation}%
which, unlike its 1D counterpart (cf. Ref. \cite{we}), does not admit exact
analytical solutions. An approximate solution can be constructed on a
truncated lattice. The simplest version of the truncation is shown in Fig.~%
\ref{sketch}, where four independent amplitudes obeying Eq. (\ref{U}) are
defined: $U_{0}$ (at the center), $U_{1}$ (in the first rhombic layer
surrounding the center), and $U_{2,3}$ (two independent amplitudes in the
second rhombic layer). The amplitudes in all other layers are neglected: $%
U_{4,5,\hdots}=0$. At $\left( m,n\right) \neq \left( 0,0\right) $, the so
truncated Eq.~(\ref{U}) yields:
\begin{equation}
U_{3}=U_{1}/\left( 2K\right) ,~U_{2}=U_{1}/K,~U_{1}=2KU_{0}/\left(
4K^{2}-5\right) ,  \label{3210}
\end{equation}%
where central amplitude $U_{0}$ is defined to be real, and the complex
coefficient is
\begin{equation}
K\equiv 2+k-i\gamma .  \label{K}
\end{equation}

The remaining nonlinear equation (\ref{eq:gl}) at the HS site, $m=n=0$,
reduces to a complex equation relating real peak intensity $U_{0}^{2}$ and
propagation constant $k$:%
\begin{equation}
iK\left( 4K^{2}-9\right) /\left( 4K^{2}-5\right) -\left( \Gamma _{1}+i\Gamma
_{2}\right) =\left( iB-E\right) U_{0}^{2},  \label{0}
\end{equation}%
which can be solved numerically. Subsequently, the amplitudes in the
surrounding layers are obtained from Eq.~(\ref{3210}).

The accuracy of the truncated-lattice analysis can be improved by including
more rhombic layers in the calculation. Although the respective version of
Eqs.~(\ref{3210}) and~(\ref{0}) become more complicated, they are still easy
to solve numerically. Table~\ref{tab1} shows numerically found $U_{0}$ and $%
k $ for the different truncation settings at two different sets of
parameters. Naturally, convergence of the solutions for the localized modes
is observed with the increase of the number of the rhombic layers.

\begin{table}[t]
\caption{Numerical solutions for different versions of the truncated lattice
at $B=1$, $E=0.1$, $\protect\gamma =0.5$, and $\Gamma _{2}=0.8$.}
\label{tab1}
\begin{center}
\begin{tabular}{|c|c|c|c|c|}
\hline
{} & \multicolumn{2}{c|}{$\Gamma_1 = 0.8574$} & \multicolumn{2}{c|}{$%
\Gamma_1 = 0.7861$} \\ \hline
{Number of rhombic layers} & $k$ & $U_0$ & $k$ & $U_0$ \\ \hline
2 & 2.2926 & 1.8020 & 1.4005 & 1.5107 \\ \hline
3 & 2.2870 & 1.8002 & 1.3633 & 1.4962 \\ \hline
4 & 2.2866 & 1.8000 & 1.3580 & 1.4942 \\ \hline
\end{tabular}%
\end{center}
\end{table}

\section{The linear-stability analysis}

\label{sec:stab}

The stability of the pinned modes found as outlined above was studied by
means of the linearization procedure~\cite{ted}. To this end, perturbed
solutions were taken as
\begin{equation}
u_{m,n}=\left[ U_{m,n}+\epsilon V_{m,n}(z)\right] e^{ikz}\;,
\label{eq:ansatz}
\end{equation}%
where $V_{m,n}(z)=X_{m,n}(z)+iY_{m,n}(z)$ is a complex-valued perturbation
with infinitesimal amplitude $\epsilon $. Substituting this, along with $%
U_{m,n}$ split into the real and imaginary parts as per Eq. (\ref{U}), into
Eq.~(\ref{eq:gl}) results in a linear system that governs the evolution of
perturbations $X_{m,n}$ and $Y_{m,n}$:
\begin{eqnarray}
\frac{dX_{m,n}}{dz} &=&-\frac{1}{2}\left(
Y_{m-1,n}+Y_{m+1,n}+Y_{m,n-1}+Y_{m,n+1}-4Y_{m,n}\right)  \notag \\
&&+kY_{m,n}-\gamma X_{m,n}+\delta _{m,0}\delta _{n,0}\left\{ \left( \Gamma
_{1}X_{m,n}-\Gamma _{2}Y_{m,n}\right) \right.  \notag \\
&&-B\left[ 2P_{m,n}Q_{m,n}X_{m,n}+\left( P_{m,n}^{2}+3Q_{m,n}^{2}\right)
Y_{m,n}\right]  \notag \\
&&-\left. E\left[ \left( 3P_{m,n}^{2}+Q_{m,n}^{2}\right)
X_{m,n}+2P_{m,n}Q_{m,n}Y_{m,n}\right] \right\} ,  \notag \\
\frac{dY_{m,n}}{dz} &=&\frac{1}{2}\left(
X_{m-1,n}+X_{m+1,n}+X_{m,n-1}+X_{m,n+1}-4X_{m,n}\right)  \notag \\
&&-kX_{m,n}-\gamma Y_{m,n}+\delta _{m,0}\delta _{n,0}\left\{ \left( \Gamma
_{2}X_{m,n}+\Gamma _{1}Y_{m,n}\right) \right.  \notag \\
&&-B\left[ \left( 3P_{m,n}^{2}+Q_{m,n}^{2}\right)
X_{m,n}+2P_{m,n}Q_{m,n}Y_{m,n}\right]  \notag \\
&&-\left. E\left[ 2P_{m,n}Q_{m,n}X_{m,n}+\left(
P_{m,n}^{2}+3Q_{m,n}^{2}\right) Y_{m,n}\right] \right\} .  \label{linear}
\end{eqnarray}%
The eigenvalue problem is obtained by substituting $X_{m,n}=\phi _{m,n}\exp
(\lambda z)$ and $Y_{m,n}=\psi _{m,n}\exp (\lambda z)$ in Eqs. (\ref{linear}%
), the underlying stationary mode $u_{m,n}(z)$ being linearly stable if all
eigenvalues $\lambda $ have $\mathrm{Re}(\lambda )\leq 0$. For the use in
the stability analysis, modal amplitudes $U_{m,n}$ were found numerically by
solving the full system~obtained from Eq. (\ref{eq:gl}) by the substitution
of expression (\ref{eq:pw}), rather than from the truncated-lattice
approximation, although the difference is very small. The numerical solution
was carried out with periodic boundary conditions, imposed onto a domain of
a size which was much larger than the width of any soliton considered below.

\begin{figure}[t]
\begin{center}
\includegraphics[width = 90mm,keepaspectratio]{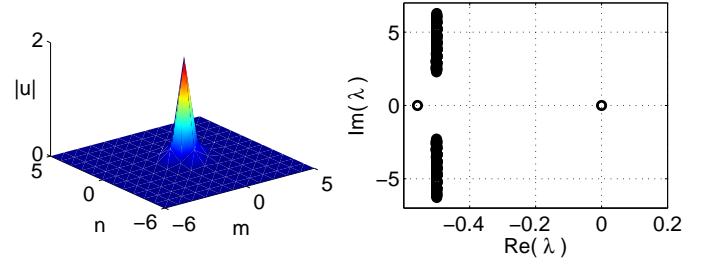}
\end{center}
\caption{Left: An example of the stable mode in the form of~(\protect\ref%
{eq:pw}) with peak amplitude $U_{0,0}=1.8$ and propagation constant $%
k=2.2865 $. Parameters are $B=1$, $E=0.1$, $\protect\gamma =0.5$, $\Gamma
_{1}=0.8574$, and $\Gamma _{2}=0.8$. Right: The eigenvalue spectrum
associated with the linearized system~(\protect\ref{linear}). All
eigenvalues have non-positive real parts.}
\label{fig2}
\end{figure}

Figure~\ref{fig2} shows a typical stable mode and its stability-eigenvalue
spectrum. The mode is peaked at the HS and symmetric about it (the shape of
this mode and its propagation constant are very close to their counterparts
produced by the truncated-lattice approximation).

\subsection{The self-focusing nonlinearity: $B=+1$}

\begin{figure}[t]
\begin{center}
\includegraphics[width = 85mm,keepaspectratio]{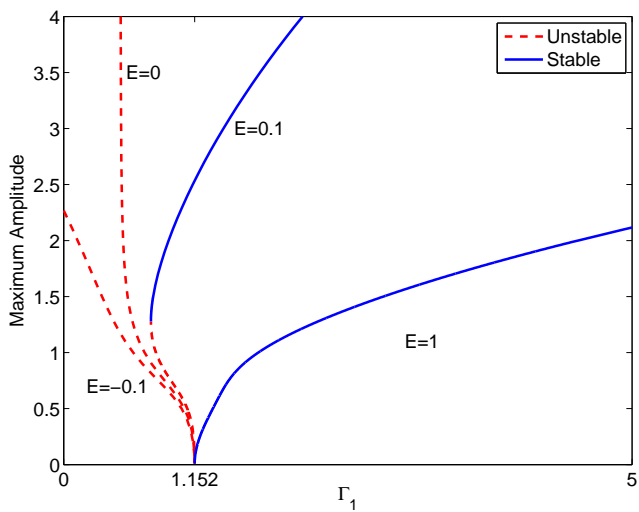}
\end{center}
\caption{(Color online) Solution branches for different values of cubic
dissipation $E$ in the self-focusing regime, $B=+1$. Here and in similar
figures below, stable and unstable branches are denoted by blue solid and
red dotted lines, respectively. Other parameters are $\protect\gamma =0.5$
and $\Gamma _{2}=0.8$.}
\label{fig3}
\end{figure}

To develop the systematic analysis, we first address the stability of pinned
modes in the self-focusing regime ($B=1$). Figure~\ref{fig3} shows solution
branches as functions of the localized linear gain $\Gamma _{1}$ at
different values of the cubic dissipation $E$. Stable solution branches can
only be found at $E>0$, i.e. in the presence of the cubic loss. For
instance, at $E=0.1$ the pinned modes with peak amplitudes $U_{0.0}>1.277$
are linearly stable, and unstable at $U_{0,0}<1.277$. Naturally, the stable
and unstable modes belong to portions of the branches where the amplitude
is, respectively, a growing or decreasing function of $\Gamma _{1}$. \ The
transition between the curves which contain the unstable segment and do not
contain it, happens, in the case displayed in Fig. \ref{fig3}, at $E=E_{%
\mathrm{cr}}\approx 0.18$.

\begin{figure}[t]
\begin{center}
\includegraphics[width = 90mm,keepaspectratio]{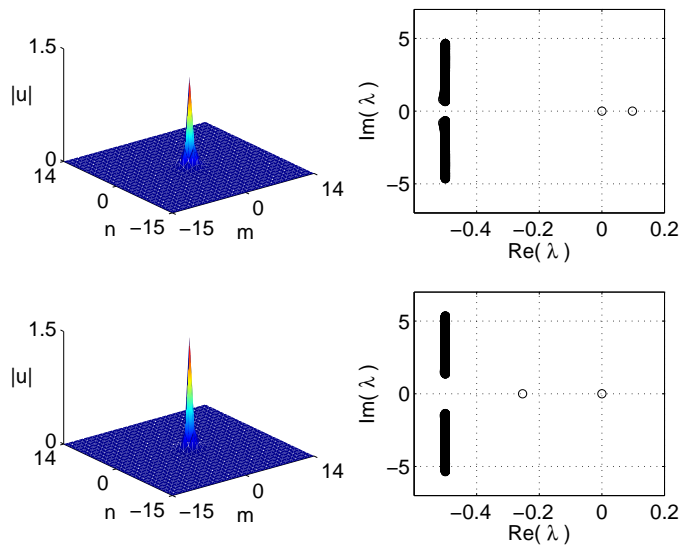}
\end{center}
\caption{Top: An unstable pinned mode with peak amplitude $U_{0,0}=1.194$
and propagation constant $k=0.6548$ at $\Gamma _{1}=0.771$ (left), and its
eigenvalue spectrum (right). Bottom: A stable pinned mode with $%
U_{0,0}=1.494 $ and $k=1.3576$ at $\Gamma _{1}=0.7861$ and its eigenvalue
spectrum. Other parameters are $B=1$, $E=0.1$, $\protect\gamma =0.5$, and $%
\Gamma _{2}=0.8$.}
\label{fig4}
\end{figure}

For the parameters considered here, stable and unstable solutions exist at $%
\Gamma _{1}\geq 0.7675$ , while at $\Gamma _{1}<0.7675$ any initial
excitation decays into the zero solution, as there is not enough energy
input to support nontrivial modes. Figure~\ref{fig4} shows some typical
examples of such stable and unstable solutions. The stability is greatly
enhanced by increasing the magnitude of the cubic loss, $E$. At $E=1$, all
the pinned modes are stable, even at very large values of $\Gamma _{1}$. On
the other hand, in the absence of the cubic dissipation ($E=0$), or in the
presence of the cubic gain ($E<0$), all the modes are unstable. These
unstable modes have essentially the same profile as in Fig.~\ref{fig4},
therefore they are not shown here.

\begin{figure}[t]
\begin{center}
\includegraphics[width = 85mm,keepaspectratio]{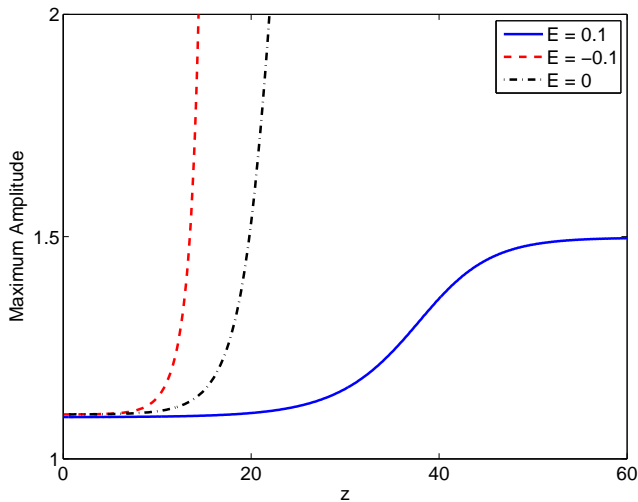}
\end{center}
\caption{The evolution of amplitudes of unstable modes at $E=0.1$, $E=-0.1$,
and $E=0$, respectively. $\Gamma _{1}$ is chosen such that the peak
amplitudes of the respective stationary solutions are $U_{0,0}\approx 1.1$.
Other parameters are $B=1$, $\protect\gamma =0.5$, and $\Gamma _{2}=0.8$.}
\label{fig4_1}
\end{figure}

Figure~\ref{fig3} shows that all the solution branches emerge from the
critical value of the linear gain, $\Gamma _{1}\approx 1.152$, which is
explained below. Furthermore, the unstable branch corresponding to $E=0$
approaches a vertical asymptote at $\Gamma _{1}=0.5$, near which the peak
amplitude increases drastically. Finally, Fig.~\ref{fig4_1} shows typical
evolution of peak amplitudes of the unstable modes corresponding to
different values of $E$, as obtained from full simulations of Eqs.~(\ref%
{eq:gl}). In the case of $E>0$, the cubic loss stabilizes the system, and
therefore the unstable mode evolves into a stable one existing at the same $%
\Gamma _{1}$, see Fig. \ref{fig3}. However, at $E\leq 0$ (the cubic gain, or
zero loss), unstable modes quickly blow up, as there are no stable branches
that might serve as attractors in this case.

\subsection{The self-defocusing nonlinearity: $B=-1$}

\begin{figure}[t]
\begin{center}
\includegraphics[width = 90mm,keepaspectratio]{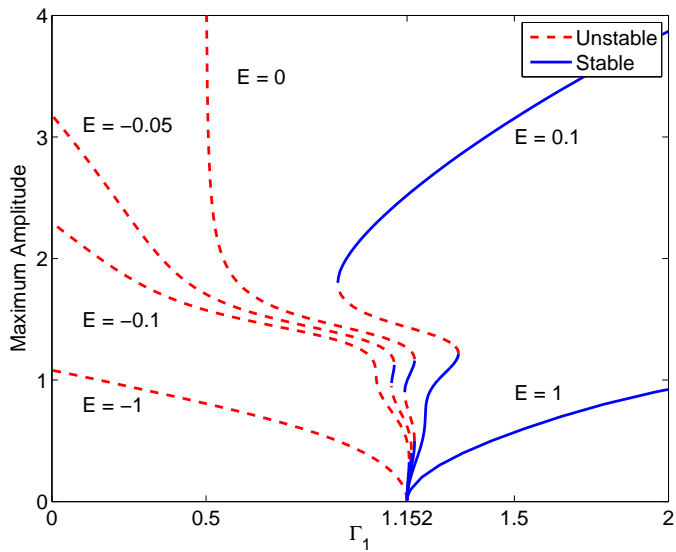}
\end{center}
\caption{(Color online) Solution branches for various values of the cubic
dissipation, $E>0$, or cubic gain, $E<0,$ in the self-defocusing regime,
with $B=-1$. Other parameters are $\protect\gamma =0.5$ and $\Gamma _{2}=0.8$%
.}
\label{fig5}
\end{figure}

\begin{figure}[t]
\begin{center}
\includegraphics[width = 90mm,keepaspectratio]{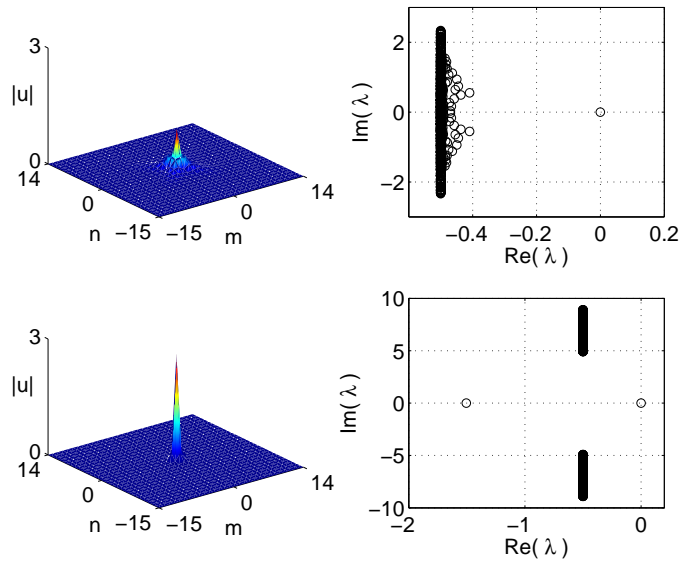}
\end{center}
\caption{Bistable solutions with amplitudes $1.04$ (top) and $2.75$
(bottom). Both solutions are obtained with $\Gamma _{1}=1.268$, $B=-1$, $%
E=0.1$, $\protect\gamma =0.5$, and $\Gamma _{2}=0.8$.}
\label{fig6}
\end{figure}

Figure~\ref{fig5} shows solution branches in the self-defocusing regime,
with $B=-1$. In the presence of a small cubic loss, the solution branch
exhibits a \textit{bistability} for a certain range of values of linear gain
$\Gamma _{1}$. For instance, at $E=0.1$, stable solutions with different
amplitudes coexist in the region of $1.152\leq \Gamma _{1}\leq 1.319$. The
two stable branches are connected by an unstable one, whose modal profile
and stability spectrum are similar to those found in the self-focusing case,
see the top panel in Fig.~\ref{fig4}, therefore they are not displayed here.
An example of the bistability is shown in Fig.~\ref{fig6}. Naturally, the
mode with the smaller peak amplitude is broader. When $E$ increases, the
unstable branch eventually gets stabilized by the strong cubic loss and
disappears from the bifurcation diagram, which happens, in the case shown in
Fig. \ref{fig6}, at $E=E_{\mathrm{cr}}\approx 0.58$. At $E>E_{\mathrm{cr}}$,
all solutions are stable, and the bistability does not take place.

\begin{figure}[t]
\begin{center}
\includegraphics[width = 90mm,keepaspectratio]{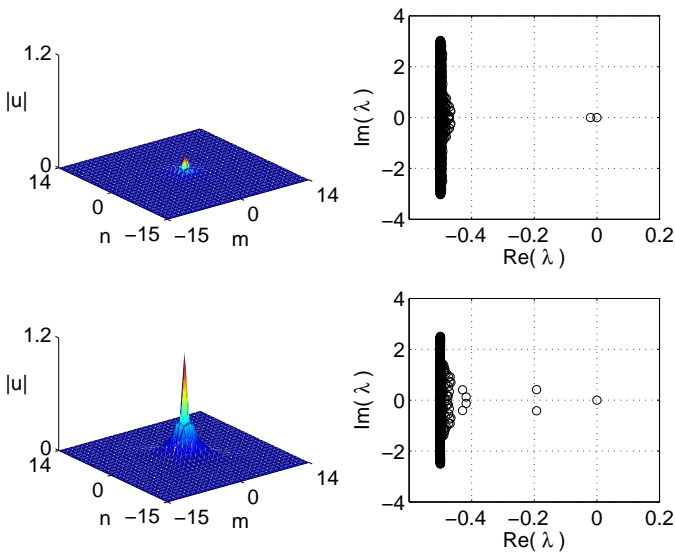}
\end{center}
\caption{(Color online) Bistable solutions with amplitudes $U_{0,0}=0.2$
(top) and $1.1$ (bottom), respectively. Both solutions are obtained with $%
\Gamma _{1}=1.159$, $B=-1$, $E=-0.01$, $\protect\gamma =0.5$, and $\Gamma
_{2}=0.8$.}
\label{fig7}
\end{figure}

\begin{figure}[t]
\begin{center}
\includegraphics[width = 85mm,keepaspectratio]{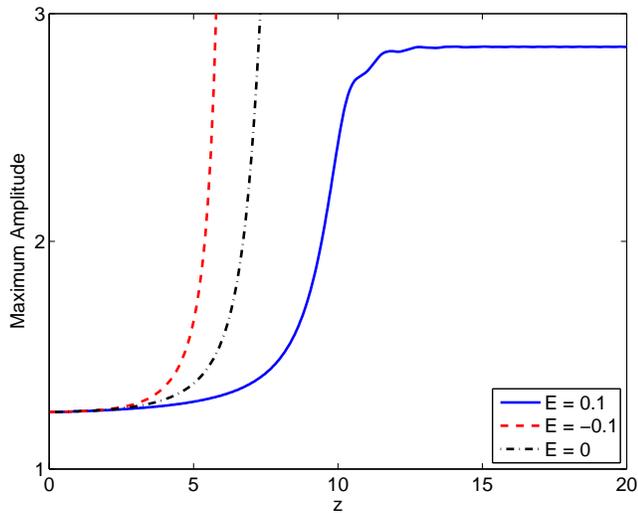}
\end{center}
\caption{(Color online) The evolution of amplitudes of unstable modes for $%
E=0.1$, $E=-0.1$, and $E=0$, respectively. $\Gamma _{1}$ is chosen such that
the initial peak amplitude is $U_{0,0}=1.25$. Other parameters are $B=-1$, $%
\protect\gamma =0.5$, and $\Gamma _{2}=0.8$.}
\label{fig7_1}
\end{figure}

A remarkable finding is that the self-defocusing gives rise to stable modes
even in the absence of the cubic loss, and in the presence of weak cubic
gain ($E\leq 0$). For small values of the cubic gain, bistability similar to
that presented in Fig.~\ref{fig7} is observed. When the cubic gain grows
(i.e. $E$ is getting more negative), the stable branch with the larger peak
amplitude disappears. All the solutions become unstable at still larger
strengths of the cubic gain. Figure~\ref{fig7_1} shows typical examples of
the evolution of the peak amplitudes of unstable modes corresponding to
different values of $E$, as produced by simulations of Eq.~(\ref{eq:gl}). In
the case of $E>0$, the initial unstable mode evolves into the closest stable
one available at the same value of $\Gamma _{1}$. As concerns unstable modes
found at $E<0$, they blow up rather than evolve into stable modes, if the
latter ones exist in the case under consideration.

\subsection{The case of $B=0$}

\begin{figure}[t]
\begin{center}
\includegraphics[width = 90mm,keepaspectratio]{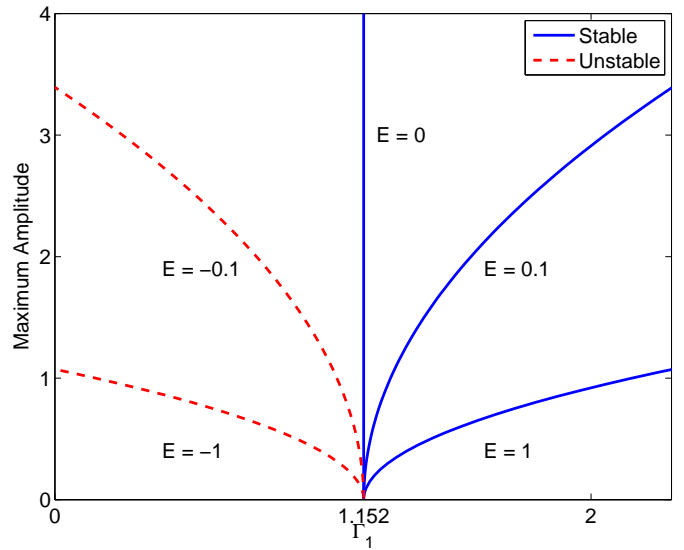}
\end{center}
\caption{(Color online) Solution branches for different values of cubic
dissipation/gain $E$ in the case of $B=0$. Other parameters are $\protect%
\gamma =0.5$ and $\Gamma _{2}=0.8$.}
\label{fig8}
\end{figure}

\begin{figure}[t]
\begin{center}
\includegraphics[width = 90mm,keepaspectratio]{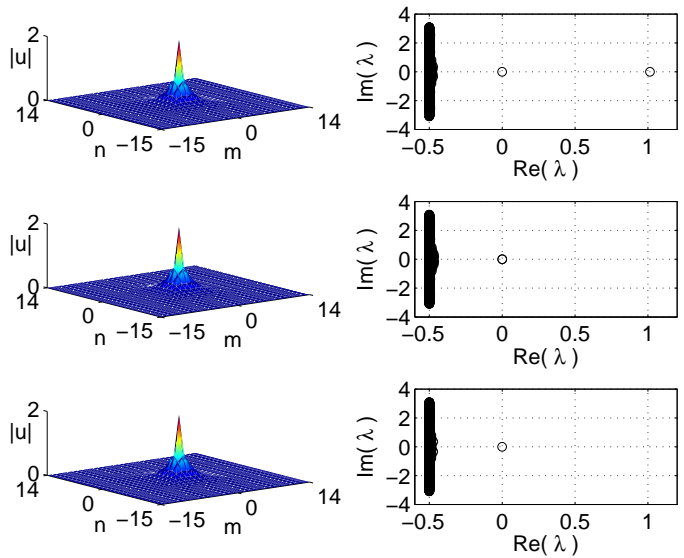}
\end{center}
\caption{Solutions with peak amplitude $U_{0,0}=2$ at different values of $%
\Gamma _{1}$ and $E$ (left) and their stability spectra (right). Top: $%
(\Gamma _{1},E)=(0.7521,-0.1)$. Middle: $(\Gamma _{1},E)=(1.1521,0)$.
Bottom: $(\Gamma _{1},E)=(1.5521,0.1)$. Other parameters are $B=0$, $\protect%
\gamma =0.5$, and $\Gamma _{2}=0.8$.}
\label{fig9}
\end{figure}

Finally, we address the stability of the pinned modes in the case of $B=0$,
when the nonlinearity is accounted for solely by the cubic loss or gain.
Figure~\ref{fig8} shows the stability of solution branches in this
situation. In particular, the solutions are always unstable in the presence
of the cubic gain ($E<0$), and always stable in the presence of the cubic
loss ($E>0$). If the cubic dissipation vanishes as well, i.e. $E=0$, Eq. (%
\ref{eq:gl}) becomes linear, admitting a single stable solution at a
specific value of the HS gain, which is $\Gamma _{1}=1.152$ for the chosen
parameters, that provides for the compensation of the background loss. This
solution obviously has an arbitrary amplitude, which explains why the
corresponding branch is vertical in Fig. \ref{fig8}. Further, Fig.~\ref{fig9}
shows examples of stable and unstable modes and their (in)stability spectra.
Unstable modes in the case of $B=0$ case evolve into the zero solution
rather than blowing up, in spite of the action of the local cubic gain, in
that case.

\section{Onset of instability of the zero solution}

\label{sec:onset}

In this section we again address the linearized version of Eq.~(\ref{eq:gl}%
), setting $B=E=0$, and study the stability of the zero solution around the
hot spot, which is an obvious condition necessary for the background
stability of pinned modes in the nonlinear system. The onset of the
background instability exactly corresponds to the existence of the
above-mentioned pinned-mode solution to the linearized equation, which
implies the equilibrium between the bulk loss ($\sim \gamma $) and local
gain ($\sim \Gamma _{1}$), in the presence of the HS's attractive potential (%
$\sim \Gamma _{2}$). Unlike the 1D counterpart of the present model \cite{we}%
, for the 2D lattice this solution of the linear equation cannot be found in
an analytical form. However, the truncated-lattice approximation presented
above in Sec.~\ref{sec:trun} provides an efficient way to study the onset of
the background instability. In particular, the onset is determined by the
critical value of $\Gamma _{1}$ that makes Eq.~(\ref{0}) with $B=E=0$
solvable. These values, found for different truncated configurations, are
summarized in Table~\ref{tab2}. As the number of the rhombic layers included
in the calculations increases, the critical value of $\Gamma _{1}$ slowly
converges to the above-mentioned numerically found value, $\Gamma
_{1}\approx 1.152$, from which the solution branches emanate in Figs.~\ref%
{fig3}, \ref{fig5}, and \ref{fig8}.

\begin{table}[t]
\caption{Critical values of $\Gamma _{1}$ for the onset of the background
instability in different truncated lattices, at $\protect\gamma =0.5$ and $%
\Gamma _{2}=0.8$.}
\label{tab2}
\begin{center}
\begin{tabular}{|c|c|}
\hline
{Number of rhombic layers} & $\Gamma_1$ \\ \hline
2 & 1.4756 \\ \hline
3 & 1.0078 \\ \hline
4 & 1.0955 \\ \hline
5 & 1.1886 \\ \hline
6 & 1.1411 \\ \hline
\end{tabular}%
\end{center}
\end{table}

\section{Dual hot spots: Symmetric and antisymmetric modes}

\label{sec:dual}

\begin{figure}[t]
\centering
\begin{tabular}{c}
\includegraphics[width = 80mm, keepaspectratio]{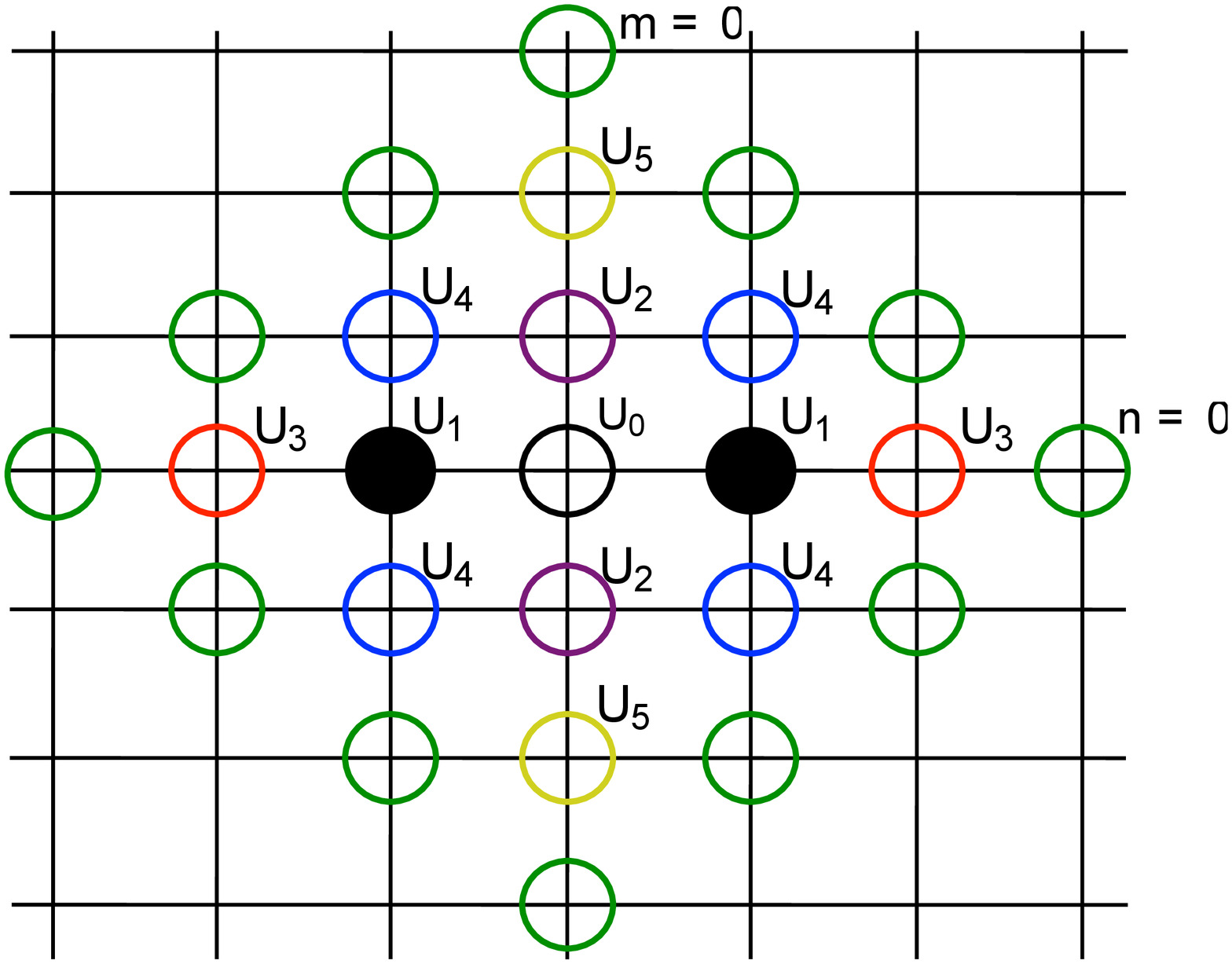} \\
\includegraphics[width = 80mm, keepaspectratio]{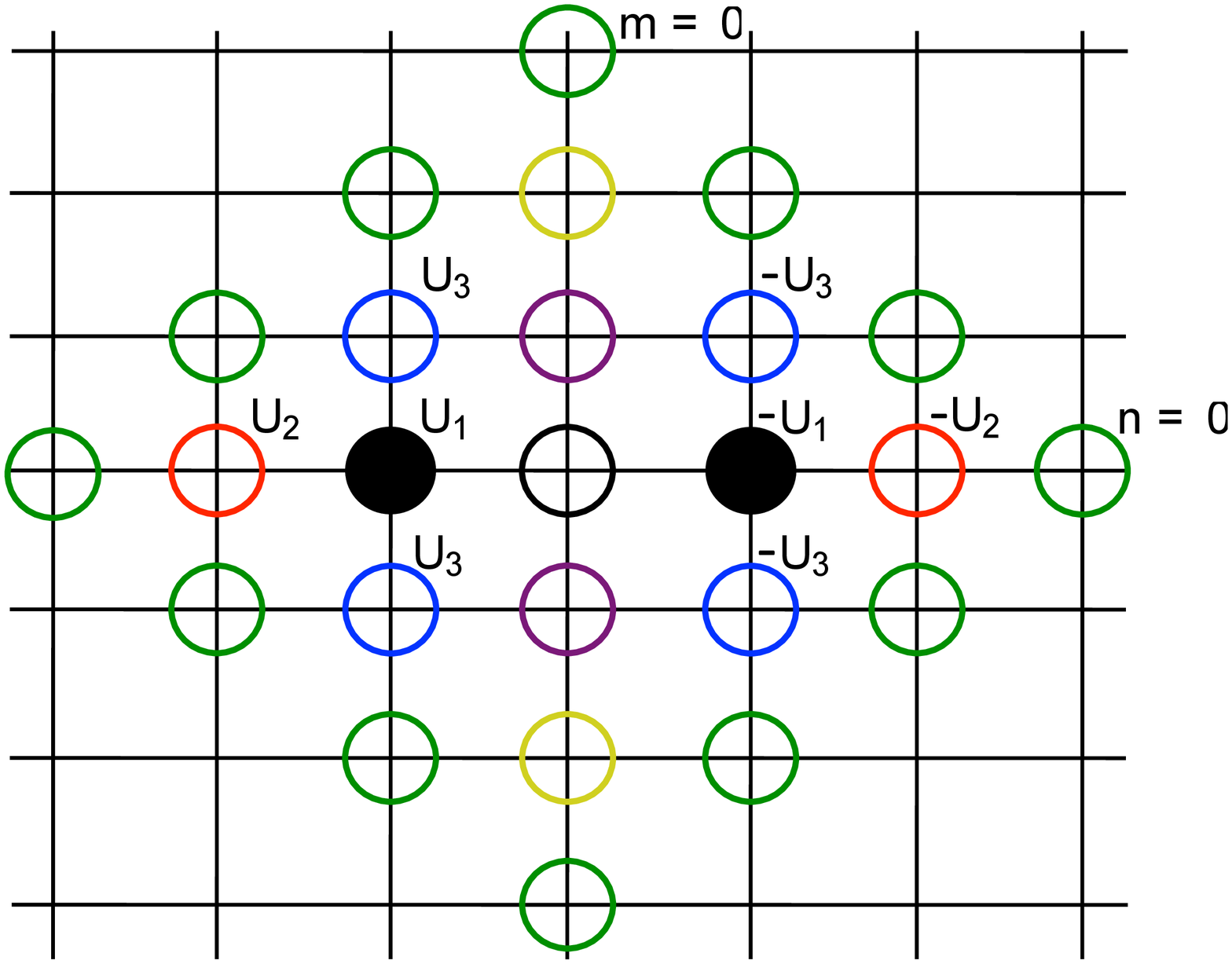}%
\end{tabular}%
\caption{(Color online) The sketch of the truncated lattice containing the
dual HSs with the single rhombic layer for symmetric (top) and antisymmetric
(bottom) modes. Here $U_{1}$ denotes the real amplitude (real) at the HSs.
In the antisymmetric mode, the amplitudes vanish at $m=0$. For both models,
the amplitudes at peripheral sites (green circles) are set to be zero.}
\label{fig10}
\end{figure}

Finally, we briefly consider pinned modes in the lattice with two identical
HSs. The truncated-lattice approximation was first used to derive simplified
equations like Eq.~(\ref{0}) to describe this configuration. To illustrate
the situation, we here place two HSs at $n=0$ and $m=\pm 1$. Pinned modes
supported by the dual HS can be naturally classified as symmetric,
antisymmetric, and asymmetric (provided that the latter species exists) \cite%
{embed1,embed2,embed3}. While amplitudes of the solutions at the two HSs may
be assumed real without the loss of generality, the symmetric and
antisymmetric modes are those with $U_{-1,0}=U_{1,0}$ and $U_{-1,0}=-U_{1,0}$%
, respectively. The simplest truncated-lattice approximations, with a single
rhombic layer surrounding the two HSs, are shown in Fig.~\ref{fig10}. For
the symmetric mode (the top panel), a calculation similar to that presented
in Sec.~\ref{sec:trun} yields equation
\begin{equation}
\frac{i\left( 16K^{4}-40K^{2}+9\right) }{4K\left( 4K^{2}-5\right) }-\left(
\Gamma _{1}+i\Gamma _{2}\right) =\left( iB-E\right) U_{1}^{2}\;,  \label{1}
\end{equation}%
with $K$ given by~Eq. (\ref{K}), whereas the antisymmetric mode (the bottom
panel) gives
\begin{equation}
\frac{i\left( 4K^{2}-3\right) }{4K}-\left( \Gamma _{1}+i\Gamma _{2}\right)
=\left( iB-E\right) U_{1}^{2}\;.  \label{2}
\end{equation}

\begin{figure}[t]
\begin{center}
\includegraphics[width = 90mm,keepaspectratio]{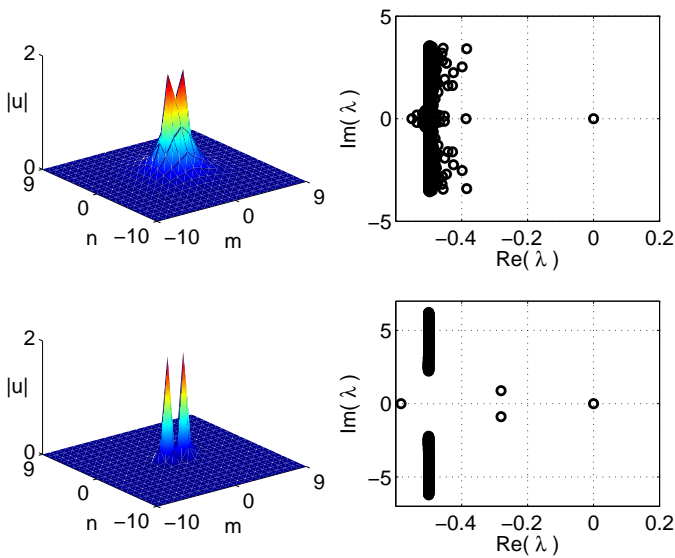}
\end{center}
\caption{(Color online) Top: A stable symmetric mode (left) with peak
amplitude $1.80$ and $k=-3.5445$ at $\Gamma _{1}=1.5445$ and $\Gamma
_{2}=-4.2224$, and its eigenvalue spectrum (right). Bottom: The same for a
stable antisymmetric mode with peak amplitude $1.8344$ and $k=2.2243$ at $%
\Gamma _{1}=0.8484$ and $\Gamma _{2}=0.8$. Other parameters are $B=1$, $%
E=0.1 $, and $\protect\gamma =0.5$. The two HSs are placed at $n=0$ and $%
m=\pm 1$.}
\label{fig11}
\end{figure}

Table~\ref{tab3} presents numerically found values of the propagation
constant, $k$, and HS amplitude, $U_{1}$, for both the symmetric and
antisymmetric modes. Figure~\ref{fig11} shows that stable pinned modes, both
symmetric or antisymmetric, can be supported by the dual HS. Peak amplitudes
of these modes (see the figure caption) are in good agreement with the
prediction of the truncated-lattice approximation.

\begin{table}[t]
\caption{Numerical solutions for the dual-HS modes in different truncated
lattices, at $B=1$, $E=0.1$, and $\protect\gamma =0.5$. For the symmetric
modes, $\Gamma _{1}=1.5445$ and $\Gamma _{2}=-4.2224$. For the antisymmetric
ones, $\Gamma _{1}=0.8484$ and $\Gamma _{2}=0.8$.}
\label{tab3}
\begin{center}
\begin{tabular}{|c|c|c|c|c|}
\hline
{} & \multicolumn{2}{c|}{Symmetric} & \multicolumn{2}{c|}{Antisymmetric} \\
\hline
{Number of rhombic layers} & $k$ & $U_1$ & $k$ & $U_1$ \\ \hline
1 & -3.4439 & 1.8903 & {2.2535} & {1.811} \\ \hline
2 & -3.6721 & 1.8114 & {2.2271} & {1.8009} \\ \hline
\end{tabular}%
\end{center}
\end{table}

\section{Conclusions}

\label{sec:con}

We have introduced the 2D discrete dynamical system based on the bulk linear
lossy lattice, into which one or two nonlinear sites with the linear gain
(HSs, \textquotedblleft hot spots") are embedded. The system can be
implemented in the form of an array of optical or plasmonic waveguides,
admitting, in particular, selective excitation of individual cores, by the
local application of the external pump to the uniformly doped array. The
analysis of localized modes pinned to the HSs was developed both
semi-analytically (using truncated lattices) and numerically. It has been
found that, in the case of the self-focusing nonlinearity at the single HS,
the pinned modes are stable (unstable) when the nonlinearity acting at the
HS contains the cubic loss (gain). On the other hand, it is worthwhile to
note that, in the case of the self-defocusing nonlinearity, the HS with an
unsaturated cubic gain supports stable modes, in a rather small parameter
region, and bistability occurs in the case of weak cubic loss. Symmetric and
antisymmetric modes pinned to dual HSs were also discussed briefly.

A challenging issue for the subsequent analysis is the possibility of the
existence of asymmetric modes supported by the symmetric dual HSs. On the
other hand, a natural extension maybe to embed one or two HSs into a
nonlinear lossy lattice.

\section{Acknowledgement}

Partial financial supported has been provided by the Hong Kong Research
Grants Council through General Research Fund contract HKU 711713E.

\end{document}